\documentclass[prb,preprint,superscriptaddress,showpacs,10pt]{revtex4-1}

\usepackage{bm}
\usepackage{graphicx}
\usepackage{amsmath}
\usepackage{hyperref}
\usepackage{epsfig}
\usepackage{epstopdf}

\usepackage[ngerman, english]{babel}

\DeclareGraphicsExtensions{.eps}

\usepackage{hyperref}
\hypersetup{%
	colorlinks=true,
	linkcolor=blue,
	urlcolor=blue,
	citecolor=blue
}

\begin{document}

\title{ Rashba semiconductor as spin Hall material: Experimental demonstration of spin pumping in wurtzite $n$-GaN:Si}

\author{R. Adhikari}
\email{rajdeep.adhikari@jku.at}
\affiliation{Institut f\"ur Halbleiter-und-Festk\"orperphysik, Johannes Kepler University, Altenbergerstr. 69, A-4040 Linz, Austria}

\author{M. Matzer}
\affiliation{Institut f\"ur Halbleiter-und-Festk\"orperphysik, Johannes Kepler University, Altenbergerstr. 69, A-4040 Linz, Austria}

\author{A. Tarazaga Mart\'{\i}n-Luengo}
\affiliation{Institut f\"ur Halbleiter-und-Festk\"orperphysik, Johannes Kepler University, Altenbergerstr. 69, A-4040 Linz, Austria}

\author{A. Bonanni}
\email{alberta.bonanni@jku.at}
\affiliation{Institut f\"ur Halbleiter-und-Festk\"orperphysik, Johannes Kepler University, Altenbergerstr. 69, A-4040 Linz, Austria}

\begin{abstract}
	Pure spin currents in semiconductors are essential for implementation in the next generation of spintronic elements. Heterostructures of III- nitride semiconductors are currently employed as central building-blocks for lighting and high-power devices. Moreover, the long relaxation times and the spin-orbit coupling (SOC) in these materials indicate them as privileged hosts for spin currents and related phenomena. Spin pumping is an efficient mechanism for the inception of spin current and its conversion into charge current in non-magnetic metals and semiconductors with Rashba SOC $via$ spin Hall effects. We report on the generation in $n$-GaN:Si\,--\,at room temperature and through spin pumping\,--\,of pure spin current, fundamental for the understanding of the spin dynamics in these non-centrosymmetric Rashba systems. We find for $n$-GaN:Si a spin Hall angle $\theta_{\mathrm{SH}}$=$3.03\times10^{-3}$, exceeding by one order of magnitude those reported for other semiconductors, pointing at III-nitrides as particularly efficient spin current generators.  
\end{abstract}

\date{\today}

\pacs{72.25.Dc, 72.25.Mk, 76.50.+g, 85.75.-d}

\maketitle


  In the emerging field of spin-orbitronics \cite{Hoffman:2015_PRApp,Kuschel:2015_NatNanotech,Zhang:2015_JAP,Manchon:2014_NatPhys} spin-orbit coupling (SOC) is employed in both magnetic and non-magnetic materials to generate, exploit and detect spin currents. Spin currents hold the key for the realization and implementation of the next generation of spin based nanoelectronic devices with properties like non-volatility, low power consumption and dissipation.  While in magnetic materials the SOC is employed to create new classes of topological objects like magnetic skyrmions or Dzyalonshinskii-Moriya domain walls \cite{Fert:2013:NatNano,Hoffman:2015_PRApp}, spin-orbitronics in non-magnetic materials mostly addresses the spin\,-to\,-charge conversion through the spin Hall effect (SHE) \cite{Sinova:2015_RMP} and the Rashba-Edelstein effect \cite{Edelstein:1990_SSC,Vignale:2016_PRB}. The concept of SHE is borrowed from the anomalous Hall effect (AHE) where, due to the relativistic SOC, asymmetric deflection of charge carriers takes place depending on their spin orientations \cite{Nagaosa:2010_RMP}. While AHE is studied in magnetic systems, SHE is mostly observed in non-magnetic ones. Based on the concept of spin dependent Mott scattering \cite{Mott:1929_PRSA}, the SHE was predicted nearly four decades ago by D'yakonov and Perel' \cite{Dyakonov:1971_JETP} and was proposed to be the effective process for producing pure spin currents in solid state systems. However, it was not until the theoretical work of Hirsch $et\,al.$ \cite{Hirsch:1999_PRL} and Zhang $et\,al.$ \cite{Zhang:2000_PRL}, that the extrinsic SHE received renewed attention. The possibility of an intrinsic mechanism was proposed by various theoretical groups \cite{Sinova:2004_PRL,Murakami:2003_Science,Murakami:2006_ASSP}. The intrinsic SHE depends only on the electronic band structure of the material. This effect arises from the nonequilibrium dynamics of the Bloch electrons as they undergo spin precession due to an induced $k$--dependent effective magnetic field, such as the Rashba field. The extrinsic SHE, on the other hand, is the mechanism in which the spins acquire transverse velocity due to SOC during scattering of electrons \cite{Vignale:2010_JSNM, Sinova:2015_RMP, Nagaosa:2010_RMP}. The extrinsic SHE is classified according to two different underlying mechanisms $viz.$ the skew scattering and the side jump mechanism. Skew scattering is the asymmetric scattering of spin, within the scattering plane, due to an effective magnetic field gradient that arises as an effect of the SOC. The scattering plane defines the spin polarization direction of the resulting spin current. The side jump mechanism is the velocity integrated over time of deflection of electrons in opposite directions by the electric fields experienced when approaching or withdrawing an impurity. This phenomenon results in an effective transverse displacement of the electrons upon multiple scattering events. At low carrier mobilities $\sim$(10$^{-2}$-10$^{2}$)\,cm$^2$V$^{-1}$s$^{-1}$, the intrinsic and extrinsic side jump mechanisms contribute mostly to the SHE, while for carrier mobilities greater than $10^{2}\,\mathrm{cm}^2\mathrm{V}^{-1}\mathrm{s}^{-1}$, extrinsic skew scattering is the dominant process \cite{Sinova:2015_RMP, Vignale:2010_JSNM}. Moreover, for mobilites exceeding $10^{3}\,\mathrm{cm}^2\mathrm{V}^{-1}\mathrm{s}^{-1}$, spin Coulomb drag becomes the dominant mechanism \cite{Vignale:2010_JSNM} and compels the spin Hall conductivity towards a saturation value. The lack of direct electrical signals proved to be a major challenge in the observation of this effect, so that the initial experimental efforts were mostly accomplished using optical means \cite{Kato:2004_Science,Wunderlich:2005_PRL,Jungwirth:2012_NatMater}.  
  
 In a series of seminal publications, Tserkovnyak, Brataas and Bauer \cite{Tserkovnyak:2002_PRL,Tserkovnyak:2002_PRB,Brataas:2002_PRB} suggested a method for obtaining pure spin current in non-magnetic (NM) metals and semiconductors with non-negligible SOC. They proposed a spin battery \cite{Brataas:2002_PRB} based on adiabatic pumping of spins from a ferromagnetic metal or insulator (FM) grown in a FM/NM bilayer configuration when the system is driven to resonance under microwave irradiation. Such a  battery leads to the dynamic generation of pure spin current in a NM with non trifling SOC $via$ ferromagnetic resonance (FMR) of the FM \cite{Ando:2014_SST}. The magnetization dynamics of ferromagnets is well described by the phenomenological Landau-Lifshitz-Gilbert (LLG) equation \cite{Tserkovnyak:2005_RMP}$\colon$ 
 
 \begin{equation} 
 \allowdisplaybreaks
 \frac{\text{d}\vec{M}\mathrm{(t)}}{\text{d}t}=-{\gamma}\vec{M}\mathrm{(t)}\times\vec{H}_\mathrm{eff}+\frac{\alpha}{M_\mathrm{s}}\vec{M}\mathrm{(t)}\times\frac{\text{d}\vec{M}\mathrm{(t)}}{\text{d}t}\label{Eq.1}
 \end{equation}
 
where $\vec{M}\mathrm{(t)}/M_\mathrm{s}$ is the unit vector of magnetization, $\gamma$ the gyromagnetic ratio and $\vec{H}_\mathrm{eff}$ the effective magnetic field expressed as  $\vec{H}_\mathrm{eff}=\vec{H}+\vec{H}_\mathrm{M}\mathrm{(t)}+\vec{h}\mathrm{(t)}$, with $\vec{H}$ the external magnetic field, $\vec{H}_\mathrm{M}\mathrm{(t)}$ the dynamic demagnetizing field and $\vec{h}_\mathrm{MW}\mathrm{(t)}$ the ac field due to microwave radiation. The magnetization $\vec{M}\mathrm{(t)}$, is expressed as a sum of the static and dynamic components, $i.e.$ $\vec{M}\mathrm{(t)}=\vec{M}\,+\,\vec{m}\mathrm{(t)}$ and the dimensionless coefficient $\alpha$ is the Gilbert damping parameter. The first term on the right hand side of Eq.\,\eqref{Eq.1} is the precession term, while the second term represents a damping component that impels the precession of magnetization $\vec{M}\mathrm{(t)}$ to spiral down to a static magnetization axis due to the Gilbert damping parameter $\alpha$. In a FM/NM hybrid bilayer, $\alpha$ is enhanced due to the transfer of spin angular momentum from the FM to the NM through a dynamical process of adiabatic spin pumping \cite{Saitoh:2006_APL,Ando:2014_SST, Tserkovnyak:2002_PRB,Pu:2015_PRL}. The spins pumped in the NM are scattered by the effective spin-orbit field $i.e.$ the Rashba field and a spin accumulation is achieved in the NM, leading to a spin current in the NM through SHE. An enhanced $\alpha$ in a FM/NM bilyer is a signature of spin pumping and a fingerprint of the generation of pure spin current in the NM. The spin current is converted into charge current, through a reciprocal process called the inverse spin Hall effect (ISHE) $via$ the relation: $\vec{J}_\mathrm{c}= {\theta _\mathrm{SH}}\vec{J}_\mathrm{s}\times {\vec\sigma}$, where $\theta _\mathrm{SH}$ is the spin Hall angle representing the efficiency of spin-to-charge conversion of a material. The charge current in the NM with SOC induces an electromotive force (emf), whose direction is perpendicular to the spin current $\vec{J}_\mathrm{s}$ and to the spin polarization vector $\vec\sigma$. Spin pumping is an efficient mechanism to convert spin current into charge current without any applied bias. Over the last few years several experimental works have been published demonstrating spin pumping in heavy metals like Pt\,[\onlinecite{Saitoh:2006_APL, Ando:2008_PRB, Ando:2011_JAP, Azevedo:2011_PRB, Boone:2015_JAP}], Ta\,[\onlinecite{Allen:2015_PRB,Avci:2014_PRB}], Pd\,[\onlinecite{Ando:2013_PRB, Boone:2015_JAP}], Au\,[\onlinecite{Heinrich:2011_PRL}], in semiconductors like GaAs [\onlinecite{Chen:2013_NatCom,Liu:2016_NatCom}], Si\,[\onlinecite{Ando:2012_NatCom, Ando:2013_PRB}], Ge\,[\onlinecite{Duschenko:2015_PRL}] and ZnO\,[\onlinecite{D'Ambrosio:2015_JJAP,Lee:APL_2014}] and recently also in 3D topological insulators [\onlinecite{Jamali:2015_NanoLett}] and ferroelectric Rashba semiconductors [\onlinecite{Rinaldi:2016_APLMater}]. Apart from ferromagnetic spin pumping from a metallic FM like permalloy (Py), it was also shown that spin currents can be realised in bilayers like Pt/YIG\,[\onlinecite{Kajiwara:2010_Nature, Sandweg:2011_PRL,Cornelissen:2015_NatPhys}] based on ferromagnetic insulators. Recent reports also point to spin pumping from paramagnets\,\cite{Shiomi:2014_PRL} and antiferromagnets \cite{Frangou:2016_PRL,Cheng:2014_PRL}. The successful generation and control of spin current in semiconductors using the mechanism of spin pumping would open wide perspectives for the integration of spin functionalities in state-of-the-art electronic and optoelectronic devices based on semiconductors like Si, Ge, III-nitrides, III-arsenides $etc$.	
  
 Amongst the conventional III-V semiconductors, GaN and its alloys AlGaN and InGaN have emerged as strategic materials for optoelectronic and electronic applications, primarily due to their tunable wide bandgap and structure induced polarization. Furthermore, transition metal doped III-nitrides have been studied extensively in the past decade as workbench magnetic semiconductors \cite{Bonanni:2011_PRB,Dietl:2015_RMP}. Another fundamental aspect, namely the presence of Rashba spin orbit coupling (RSOC) \cite{Rashba:1960_Fiz,Rashba:1965_SovPhys}, was recently demonstrated in degenerate wurtzite (wz) $n$-GaN:Si [\onlinecite{Stefanowicz:2014_PRB}]. Using magnetotransport measurements it was shown that the Rashba parameter, $\alpha_\mathrm{R}$, linear in $k$ and accounting for the spin splitting of the conduction band in wz-GaN, has a quantitative value measured for $n$-GaN:Si to be $\sim$ ($ 4.5\pm1.0$)\,$\mathrm{meV}{\cdot}{\mathrm{\AA}}$ $i.e.$ the same magnitude found for a 2-dimensional electron gas (2DEG) formed at the AlGaN/GaN interface \cite{Stefanowicz:2014_PRB}. In this work it was shown that in polar wz-GaN, the inversion asymmetry associated with the wurtzite crystal structure dominates over the interfacial electric field in the conduction band of GaN. The RSOC in degenerate $n$-GaN:Si makes this material a spin Hall system for prospective spin-orbitronic applications based on III-nitride semiconductors. Here we report on the generation of pure spin currents in $n$-GaN:Si using an adiabatic spin pumping technique and we estimate the spin Hall angle for $n$-GaN:Si. The evaluation of the spin-to-charge conversion efficiency in $n$-GaN:Si opens up the possibility to design and implement high performance spin based III-nitride nanoelectronic and optoelectronic devices.

  
 \section{Experimental details}

 \begin{figure*}[ht]
 	\centering
 	\includegraphics[width=14.5 cm]{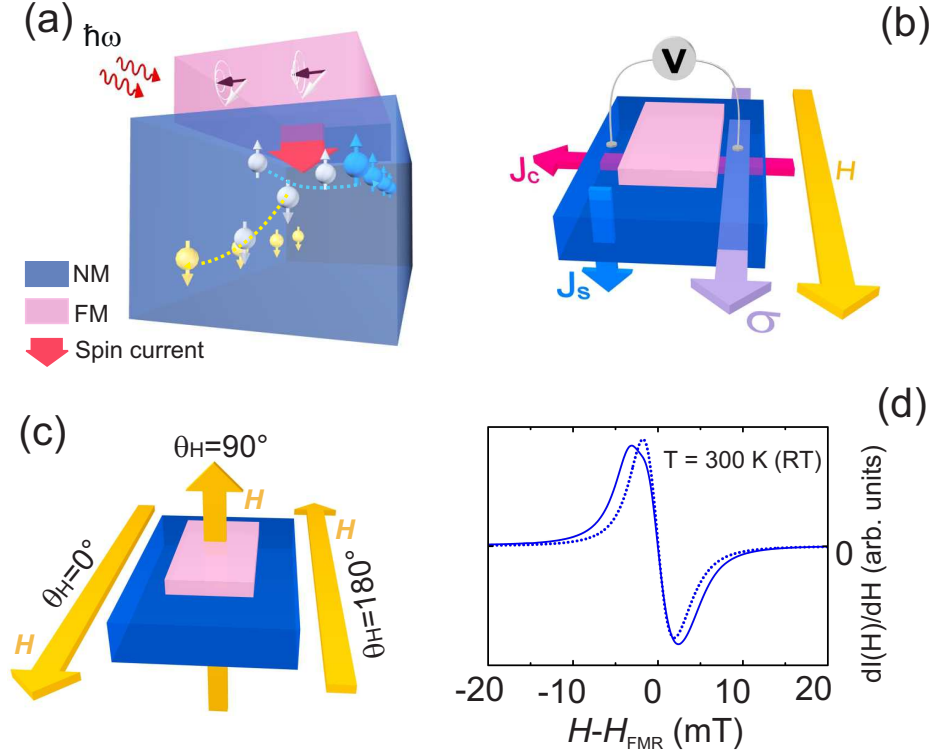}
 	\caption {(a) Spin pumping mechanism; (b) sample structure and schematic illustration of the physical quantities describing the spin pumping and spin current generation and detection through the Ohmic contacts on $n$-GaN:Si; (c) orientations of the applied magnetic field w.r.t. the sample surface  and (d) $\vec{H}$ dependence of the FMR signals dI(H)/dH measured at T = 300 K for Py/$c$-Al$_{2}$O$_{3}$ and Py/$n$-GaN:Si bilayer at $\theta_\mathrm{H}=0^\circ$.
 		\label{fig:fig1} } 	
 \end{figure*}
 
 The experiments have been performed on a 1.9~$\mu\mathrm{m}$ thick degenerately doped $n$-GaN:Si film with electron concentration ($1.2\times$\,10$^{19}$)\,cm$^{-3}$ and carrier mobility, $\mu\sim\,180\,\mathrm{cm}^2\mathrm{V}^{-1}\mathrm{s}^{-1}$, grown by metal organic vapour phase epitaxy (MOVPE) on $c$-$\mathrm{Al}_2\mathrm{O}_3$. A 10 nm permalloy (Py = $\mathrm{Ni}_{80}\mathrm{Fe}_{20}$) film is used as  FM layer and source of spins. The Py film is passivated with a 6 nm $\mathrm{AlO}_\mathrm{x}$ film  which protects the FM from oxidation in order to avoid the detrimental effects of oxidised Py on the spin pumping efficiency \cite{Kim:2014_CAP}. The Py/$n$-GaN:Si bilayer is driven to resonance conditions at room temperature under an X-band microwave excitation of ($9.44\pm0.01$)  GHz with an external magnetic field in a Bruker Elexsys E580 electron paramagnetic resonance spectrometer. The permalloy\,--\,being a soft FM\,--\,has a small magnetocrystalline anisotropy, so that $\vec{M}\mathrm{(t)}$ in Py is aligned along the film plane when an in-plane magnetic field $\vec{H}$ is applied. Ohmic contacts to $n$-GaN:Si are fabricated by e-beam evaporation of Ti/Au/Al/Ti/Au as a metallic stack. 
 
 The quartz sample holder for the FMR measurements is provided with two high conducting copper wires and connected to a Keithley 2700 DMM for measuring the generated dc voltage. The dc voltage due to inverse spin Hall-, thermal- and galvanomagnetic-effects is measured and the value of the component due to inverse spin Hall effect is employed to calculate the spin Hall angle in $n$-GaN:Si. Control experiments are also performed on contacted $n$-GaN:Si without Py, on Py/$c$-$\mathrm{Al}_2\mathrm{O}_3$, on bare $c$-$\mathrm{Al}_2\mathrm{O}_3$ substrates and on the wired sample holder solely. However, no measurable voltage or FMR absorption have been detected from these test samples and control experiments,  ruling out experimental artefacts affecting the observed results. 
 
  \section{Detection of spin Hall effect in $n$-G\lowercase{a}N:S\lowercase{i}} 
    
 A schematic representation of the spin pumping mechanism and of the geometry employed for the detection of the generated spin current in the NM are shown in Figs.\,\ref{fig:fig1}(a) and \ref{fig:fig1}(b), respectively.  The sketches depict the magnetization precession in the FM for an applied magnetic field $\vec{H}$ at resonance condition under microwave excitation, leading to the pumping of spin angular momentum and subsequent generation of spin and charge currents in the NM through the SHE and ISHE. Measurements are carried out with the magnetic field applied in the in-plane and out-of-plane configuration, respectively.
 
 Spin pumping is quenched for a perpendicular magnetic field and the angle dependent measurement of the emf is essential to rule out secondary effects or experimental artefacts. The direction of the applied magnetic field w.r.t the sample plane is indicated in Fig.\,\ref{fig:fig1}(c), while the first derivative of the FMR signal for Py/$c$-Al$_{2}$O$_{3}$ and Py/$n$-GaN:Si/$c$-Al$_{2}$O$_{3}$ is provided in Fig.\,\ref{fig:fig1}(d). The dotted and solid lines represent the FMR line-shapes for Py/$c$-Al$_{2}$O$_{3}$ and for the Py/$n$-GaN:Si bilayer, respectively. The broadening of the FMR signal for the Py/GaN:Si bilayer w.r.t. the reference Py/$c$-Al$_{2}$O$_{3}$ layer is the evidence of adiabatic spin pumping from the ferromagnetic Py into the Rashba semiconductor $n$-GaN:Si under resonance conditions \cite{Saitoh:2006_APL,Ando:2014_SST,Duschenko:2015_PRL}. The FMR signal and the electric potential difference between the electrodes attached to the $n$-GaN:Si layer are measured to detect the ISHE.  
 
 The dynamics of the magnetization $\vec{M}\mathrm{(t)}$ in Py under an effective magnetic field $\vec{H}_\mathrm{eff}$ is described by the LLG Eq.\,\eqref{Eq.1}. In FMR regime, the spin pumping driven by dynamical exchange interaction, pumps into $n$-GaN:Si pure spin current, which gets converted into charge current $via$ ISHE, according to the relation $\vec{J}_\mathrm{c}= {\theta _\mathrm{SH}}\vec{J}_\mathrm{s}\times {\vec{\sigma}}$, as discussed earlier. The charge current in the Rashba semiconductor leads to an emf proportional to the generated spin current and whose amplitude is proportional to the microwave absorption and maximized at the resonance field $H_\mathrm{FMR}$. 
	
 \begin{figure}[ht]
 	\centering
 	\includegraphics[width=9.0 cm]{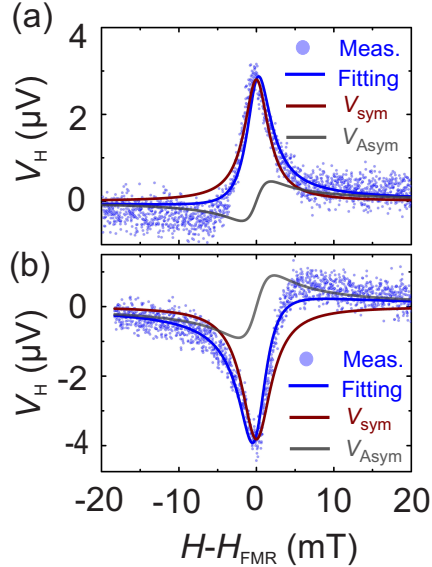}
 	\caption{Voltage across the electrodes on $n$:GaN:Si layer for in-plane magnetic field orientations at (a) $\mathrm{\theta}_\mathrm{H}=0^\circ$ and (b): $\mathrm{\theta}_\mathrm{H}=180^\circ$ for $P_\mathrm{MW} = 200\, \mathrm{mW}$. Dots: experimental data; solid lines: fitting according to Eq.\,\eqref{Eq.2}.
 		\label{fig:fig2}}	
 \end{figure}

 The emf $V_\mathrm{H}$ generated in the NM $n$-GaN:Si is detected simultaneously with the FMR and at the resonance field a peak is observed in the measured voltage. The  experimental emf is plotted as a function of the applied magnetic field $\vec{H}$ and reported in Figs.\,\ref{fig:fig2}(a) and \ref{fig:fig2}(b) respectively for $\theta_\mathrm{H}=0^\circ$ and $\theta_\mathrm{H}=180^\circ$. The angles $0^\circ$ and $180^\circ$ define the directions of the applied in-plane magnetic field in accordance to the geometry provided in Fig.\,\ref{fig:fig1}(c). Since the Rashba-Edelstein effect is observed in 2DEGs and in materials with giant Rashba SOC \cite{RojasSanchez:2013_NatCom}, and given that the system treated here belongs to neither of these two classes, the observed emf is attributed to the interplay between SHE and ISHE. The experimental emf $V_\mathrm{H}$ is a superposition of the voltage due to ISHE in the $n$-GaN:Si layer \cite{Saitoh:2006_APL,Ando:2011_NatMater} and of spurious voltages originating from galvanomagnetic effects like ordinary Hall effect (OHE) in $n$-GaN:Si, anomalous Hall effect (AHE), planar Hall effect  (PHE) in the Py layer and thermal heating effects due to microwave irradiation \cite{Juretschke:1960_JAP}.  The voltage originating from these spurious mechanisms can be separated from the one due to ISHE by fitting the experimental data with the function \cite{Saitoh:2006_APL}$\colon$ 

 \begin{equation}  
\allowdisplaybreaks
V_\mathrm{H}=V_\mathrm{Sym}\frac{\varGamma^{2}}{(H-H_\mathrm{FMR})^2+\varGamma^{2}}\,  + V_\mathrm{Asym}\frac{-2\varGamma(H-H_\mathrm{FMR})}{(H-H_\mathrm{FMR})^2+\varGamma^2}\label{Eq.2} 
 \end{equation}

 where $\varGamma$ is the half line width of the FMR line-shape and $H_\mathrm{FMR}$ is the resonance field. The symmetric part of the total voltage function, $V_\mathrm{Sym}\varGamma^{2}/\left[(H-H_\mathrm{FMR})^2+\varGamma^{2}\right]$ with an absorption line-shape is  the contribution of the ISHE voltage, $V_\mathrm{ISHE}$, developed in $n$-GaN:Si due to the adiabatic spin pumping and to heating effects. On the other hand, the asymmetric part of the function, $V_\mathrm{Asym}\left[-2\varGamma(H-H_\mathrm{FMR})\right]/\left[(H-H_\mathrm{FMR})^2+\varGamma^2\right]$ with a dysonian dispersion line-shape is a consequence of the contributions from AHE and OHE, as discussed earlier. The Hall voltages $V_\mathrm{Asym}$ change sign across $H_\mathrm{FMR}$, while $V_\mathrm{ISHE}$\,--\,being proportional to the integrated microwave absorption intensity\,--\,is symmetric across $H_\mathrm{FMR}$, as expected from the fundamental spin pumping model \cite{Tserkovnyak:2005_RMP,Saitoh:2006_APL}. A fitting of the measured emf at a microwave power of 200 mW with Eq.\,\eqref{Eq.2} for ${\theta}_\mathrm{H}=0^\circ$ yields $V_\mathrm{Sym} = 2.80\, {\mu}V$ and $V_\mathrm{Asym} = -0.453\,{\mu}V$. The ratio $V_\mathrm{Sym}$/$V_\mathrm{Asym}$\,$\sim$\,6 indicates that the major contribution to the measured voltage is provided by the efficient conversion of spin-to-charge current due to an interplay of spin pumping and direct and inverse spin Hall effects. The spurious heating effects from the microwave irradiation can be further eliminated by averaging the symmetric voltage for parallel and antiparallel orientation of the applied field $H$ according to$\colon$ 

\begin{equation} 
\allowdisplaybreaks
V_\mathrm{ISHE}(\theta_\mathrm{H})=\frac{V_\mathrm{Sym}(\theta_\mathrm{H})-V_\mathrm{Sym}(\theta_\mathrm{H}+180^\circ)}{2}\label{Eq.3}
\end{equation}

This approach to the treatment of the data is justified, since $V_\mathrm{ISHE}$ changes sign upon reversal of the magnetic field owing to a change of sign of the spin polarization vector $\vec{\sigma}$, not occurring for the voltage due to the microwave heating effects.  Thus, for spin pumping experiments, the measured emf $V_\mathrm{H}$ includes contributions from both symmetric and asymmetric voltages, and the actual ISHE voltage $V_\mathrm{ISHE}$ is estimated by a proper treatment of the measured data. It was reported \cite{Azevedo:2011_PRB,Chen:2013_NatCom,Chen:2014_APEX,Lustikova:2015_PRB}, that the symmetric voltage can also include contributions from PHE and galvanomagnetic effects in the metallic FM layer due to the electric and magnetic field components of the microwave. However, the angular dependence of the emf w.r.t. the external magnetic field showed that the contribution of the ISHE to the symmetric voltages at least five times greater than the ones from the PHE \cite{Lustikova:2015_PRB}. In the above mentioned works, the thickness of the NM layer was of the order of the spin diffusion length  $\lambda_\mathrm{N}$, while the thickness of $n$-GaN:Si studied here is  $1.9\,\mu\mathrm{m}$, $i.e.$ much greater than $\lambda_\mathrm{N}\sim$\,80\,nm in $n$-GaN. The fact that the thickness of the NM is orders of magnitude greater than $\lambda_\mathrm{N}$ suppresses the backflow spin current into the metallic FM, further reducing the contributions to the emf from galvanomagnetic effects. Careful identification and estimation of the various components of the generated emf in the NM layer in a spin pumping experiment is therefore essential for the estimation of the spin-to-charge conversion efficiency of a material.  Here, for the in-plane magnetic field, after eliminating the heating effects, a $V_\mathrm{ISHE} = 3.25\,{\mu}V$ is obtained and is exploited for the quantitative evaluation of the spin Hall angle for $n$-GaN:Si. 
 
\begin{figure*}[ht]
	\centering
	\includegraphics[width=15 cm]{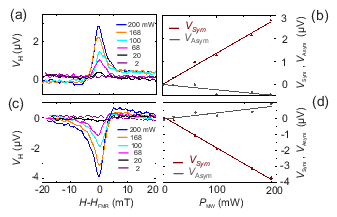}
	\caption{ \label{fig:fig3} Dependence of the measured voltage on the applied microwave power for in-plane magnetic field orientations at (a) $\theta_\mathrm{H}=0^\circ$ and (c) $\theta_\mathrm{H}=180^\circ$. Dependence of the $V_\mathrm{Sym}$ and $V_\mathrm{Asym}$ for different microwave powers for (b) $\theta_\mathrm{H}=0^\circ$ and (d) $\theta_\mathrm{H}=180^\circ$. }	
\end{figure*}

 The dependence of the measured emf on the applied microwave power under FMR conditions  is shown in Figs.\,\ref{fig:fig3}(a) and \ref{fig:fig3}(c) for orientations of the magnetic field $\theta_\mathrm{H}=0^\circ$ and $\theta_\mathrm{H}=180^\circ$, respectively. For a microwave power of 200 mW, the measured voltages at FMR conditions are +2.849 $\mu\mathrm{V}$ for $\theta_\mathrm{H}=0^\circ$ and -3.840 $\mu\mathrm{V}$ for $\theta_\mathrm{H}=180^\circ$. It is to be noted here that the voltages reported in Figs.\,\ref{fig:fig3}(a) and \ref{fig:fig3}(c) contain both the symmetric and asymmetric contributions. After separating the symmetric and asymmetric voltage components, the symmetric part of the voltage, $V_\mathrm{Sym}$ is plotted as a function of $P_\mathrm{MW}$ and shown in Figs.\,\ref{fig:fig3}(b) and \ref{fig:fig3}(d) for $\theta_\mathrm{H}=0^\circ$ and $\theta_\mathrm{H}=180^\circ$, respectively. Now, being $V_\mathrm{ISHE}$ proportional to the square of the microwave magnetic field ($h_\mathrm{MW}$), it is expected to be linearly proportional to the microwave power $P_\mathrm{MW}$\,[\onlinecite{Duschenko:2015_PRL,Ando:2011_NatMater}]. This is validated in Figs.\,\ref{fig:fig3}(b) and \ref{fig:fig3}(d) for $\theta_\mathrm{H}=0^\circ$ and $\theta_\mathrm{H}=180^\circ$, respectively. The reversal of $\vec{H}$ causes $\vec{\sigma}$ to change sign, which in turn induces the change in sign of the ISHE electric field $\vec{E}_\mathrm{ISHE}$, as evidenced in Figs.\,\ref{fig:fig3}(a) and \ref{fig:fig3}(c). It can be concluded that the fundamental relation for the ISHE, $\vec{J}_\mathrm{c}= \theta_\mathrm{SH}\vec{J}_\mathrm{s}\times \vec{\sigma}$ operates in the studied Py/$n$-GaN:Si bilayer system. For an applied magnetic field perpendicular to the sample plane, $i.e.$ $\theta_\mathrm{H}=90^\circ$, the amplitude of $V_\mathrm{ISHE}$ is quenched even though a FMR signal for the Py is detected at the resonance field of 1245 mT. This result provides the necessary and sufficient confirmation of spin pumping through SHE and ISHE in $n$-GaN:Si.  

\section{Estimation of spin Hall angle in $n$-G\lowercase{a}N:S\lowercase{i}} 

The magnetization precession described by the LLG Eq.\,\eqref{Eq.1} drives the spin pumping in the Py/$n$-GaN:Si film. Within the fundamental model of spin pumping proposed by Tserkovnyak $et\,al.$ \cite{Tserkovnyak:2002_PRL}, the dc component of the generated spin current density $j_s^0$ at the interface between Py and $n$-GaN:Si is:
  
\begin{equation} 
\allowdisplaybreaks
j_s^0=\frac{\omega}{2\pi}\int_{0}^{2\pi} \frac{\hbar}{4\pi}g_r^{\uparrow\downarrow}\frac{1}{M_{s}^2}\left[\vec{M}\mathrm(t)\times\frac{\text{d}\vec{M}\mathrm(t)}{\text{d}t}\right]_{z}dt\label{Eq.4}
\end{equation}

 where $\hbar$ and $g_r^{\uparrow\downarrow}$ are the Dirac constant and real part of the spin mixing conductance, while $\left[\vec{M}(t)\times\text{d}\vec{M}(t)/\text{d}t\right]_{z}$ is the $z$ component of $\left[\vec{M}(t)\times\text{d}\vec{M}(t)/\text{d}t\right]$. Now, $g_r^{\uparrow\downarrow}$ is proportional to the reflection and transmission coefficients of the majority and minority spins of the NM electrons, which in turn depend on the transparency of the FM/NM interface \cite{Zhang:2015_NatPhys,Pai:2015_PRB}. Thus, a transparent interface $i.e.$ one with negligible spin scattering potential centres, would enhance $g_r^{\uparrow\downarrow}$, by augmenting the spin current density in the NM.  The resonance condition obtained as a solution of the LLG equation is given by:

\begin{equation}
\allowdisplaybreaks
\left(\frac{\omega}{\gamma}\right)^2=H_\mathrm{FMR}\left(H_\mathrm{FMR}+4{\pi}M_\mathrm{s}\right)\label{Eq.5}
\end{equation}

where $H_\mathrm{FMR}$ is the resonance field, $\omega$ the frequency of the microwave radiation and $\gamma$ is the gyromagnetic ratio. Using Eqs.\,\eqref{Eq.4} and \eqref{Eq.5} and the dynamic components of the magnetization $\vec{M}\mathrm{(t)}$ obtained as a solution of Eq.\,\eqref{Eq.1} with the components of $\vec{H}_\mathrm{eff}$ as discussed above \cite{Ando:2008_PRB}, the spin current density $j_s^0$ is:

\begin{equation}
\allowdisplaybreaks
j_s^0=\frac{g_r^{\uparrow\downarrow}\gamma^{2}h_{MW}^2\hbar\left[4{\pi}M_{s}\gamma+\sqrt{{(4{\pi}M_{s}})^{2}\gamma^2)+4\omega^{2}}\right]}{8{\pi}\alpha^{2}\left[(4{\pi}M_{s})^2\gamma^2+4\omega^2\right]}\label{Eq.6}
\end{equation}

The spin Hall angle $\theta_\mathrm{SH}$ is related to the spin current $j_s^0$ and to the inverse spin Hall voltage $V_\mathrm{ISHE}$ as:

\begin{equation}
\allowdisplaybreaks
\theta_{\mathrm{SH}}=\left(\frac{\hbar}{2e}\right)\frac{\textit{V}_\mathrm{ISHE}\left(d_{N}\sigma_{N}+d_{F}\sigma_{F}\right)}{w\sigma_{N}\tanh\left(\frac{d_{N}}{2\lambda_{N}}\right)j_s^0}\label{Eq.7}
\end{equation}
 
 where $d_\mathrm{N}$, $\sigma_\mathrm{N}$ and $\lambda_\mathrm{N}$ are the thickness, conductivity and spin diffusion length of the $n$-GaN:Si layer, respectively, while $d_\mathrm{F}$ and $\sigma_\mathrm{F}$ are the thickness and conductivity of the ferromagnet Py. The real part of the spin mixing conductance $g_r^{\uparrow\downarrow}$ is given by \cite{Tserkovnyak:2005_RMP, Ando:2008_PRB, Ando:2011_JAP, Deorani:2013_APL}:
 
 \begin{equation}
 \allowdisplaybreaks
 g_r^{\uparrow\downarrow}= \frac{4{\pi}M_{s}\sqrt{3}{\gamma}{d}_\mathrm{F}}{2g\mu_{B}\mu_{0}\omega}\left(W_\mathrm{FM/NM}-W_\mathrm{NM}\right)\label{Eq.8}
 \end{equation}

 where $g$ and $\mu_\mathrm{B}$ are the Land\'{e} $g$ factor and the Bohr magneton, $W_\mathrm{FM/NM}$ and $W_\mathrm{NM}$ the spectral width of the Py/GaN:Si and Py layers, as shown in Fig.\,\ref{fig:fig1}(d). With $g = 2.12$, $4{\pi}{M}_\mathrm{s} = 0.938\,\mathrm{T}$, ${d}_\mathrm{F}$= ($1\times10^{-8}$)\,m, ${\mu}_B$=($9.27\times10^{-24}$)\,JT$^{-1}$, $\omega$=($5.931\times10^{10}$)\,s$^{-1}$ and $\mu_{0}$=($4{\pi}\times10^{-7}$)\,H/m, we calculate the spin mixing conductance $g_r^{\uparrow\downarrow}$ to be ($1.38\times10^{18}$)\,m$^{-2}$. Using Eqs.\,\eqref{Eq.6} and \eqref{Eq.7} and with the parameters $\gamma$=$(1.86\times10^{11}$)\,T$^{-1}$s$^{-1}$, ${h}_\mathrm{MW} = 0.15\, \mathrm{mT}$, $\hbar$=($1.054\times10^{-34}$)\,Js, $\lambda_\mathrm{N}=80\,\mathrm{nm}$, ${d}_\mathrm{N}=1.9\,{\mu}\mathrm{m}$, $w= 3\,\mathrm{mm}$, $d_\mathrm{F}=10\,\mathrm{nm}$, $\sigma_\mathrm{N}$=($3.5587\times10^{4}$)\,$\Omega^{-1}$m$^{-1}$, $\sigma_{F}$=($1.6\times10^{6}$)\,$\Omega^{-1}$m$^{-1}$, $j_{s}^{0}$=($1.2254\times10^{-10}$)\,Jm$^{-2}$ and $V_\mathrm{ISHE}=3.25\,{\mu}\mathrm{V}$ we find the spin Hall angle for $n$-GaN:Si to be $\theta_\mathrm{SH}=3.03\times10^{-3}$, at least one order of magnitude higher than those reported for Si [\onlinecite{Ando:2012_NatCom}], Ge[\onlinecite{Koike:2013_APE}], ZnO [\onlinecite{Lee:APL_2014}] and $n$-GaAs [\onlinecite{Ehlert:2012_PRB,Sinova:2015_RMP}]. The spin Hall angle $\theta_\mathrm{SH}$ depends on both the side jump and the skew scattering mechanisms. Theoretically\cite{Ando:2011_JAP} the side jump contribution to $\theta_\mathrm{SH}$ is given by $(3/8)^{1/2}({k}_\mathrm{F}{\lambda}_\mathrm{N})^{-1}$, which corresponds to $1.08\times10^{-2}$ in the case under consideration . The theoretical value is at least one order of magnitude higher than the one obtained experimentally. Defining the evolution of SHE in terms of mobility of the system in question according to Vignale $et\,al.$ \cite{Vignale:2010_JSNM}, with $\mu\sim\,180\,\mathrm{cm}^2\mathrm{V}^{-1}\mathrm{s}^{-1}$, $n$-GaN:Si falls in the limit of clean-ultraclean regime, which is dominated by the skew scattering. Considering the above mentioned overestimation of the magnitude of the side jump contribution and the consequent discrepancy between the theoretical and experimental values of $\theta_{\mathrm{SH}}$, one can infer that SHE in $n$-GaN:Si is dominated by skew scattering.  

 \begin{figure}[ht]
 	\centering
 	\includegraphics[width=6.5 cm]{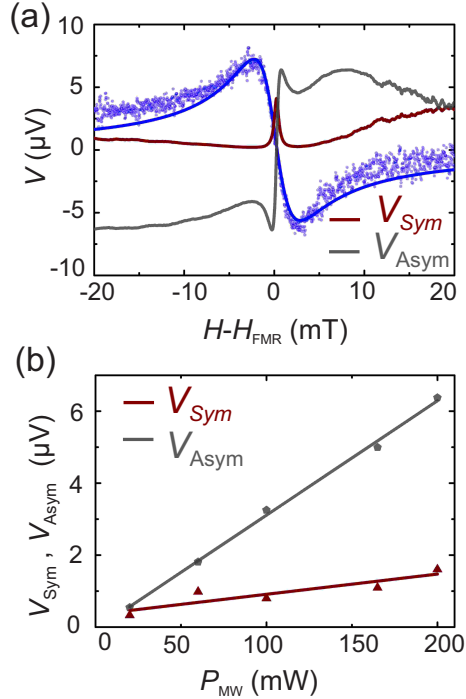}
 		\caption{(a) Voltage across the electrodes in the $150\,\mathrm{nm}$ $n$-GaN:Si control sample under a microwave excitation of power $200\,\mathrm{mW}$. Dots: experimental data; solid lines: fitting according to Eq.\,\eqref{Eq.2}. (b) Measured $V_\mathrm{ISHE}$ and $V_\mathrm{Asym}$ as a function of the microwave power. Lines: linear fits.}\label{fig:fig4}
 \end{figure}
 
The experiments discussed above for the $1.9\,\mu\mathrm{m}$ thick $n$-GaN:Si layer have been carried out also on a $150\,\mathrm{nm}$ thick  $n$-GaN:Si film. Like in the case of the $1.9\,\mu\mathrm{m}$ $n$-GaN:Si sample, an $\mathrm{AlO}_\mathrm{x}$ passivated 10 nm Py film is used as the source of spin. The spin Hall voltage measured in the 150 nm thick $n$-GaN:Si is reported in Fig.\,\ref{fig:fig4}. Here, the intensity of the asymmetric component of the voltage is more than one order of magnitude higher than the one observed for the $1.9\,{\mu}\mathrm{m}$ $n$-GaN:Si sample, as shown in Figs.\,\ref{fig:fig4}(a) and \ref{fig:fig4}(b). For a NM layer thinner than its spin diffusion length $\lambda_\mathrm{N}$, the spin backflow current $I_\mathrm{bf}$ from the NM to FM exceeds the forward spin current due to spin pumping $I_\mathrm{sp}$ [\onlinecite{Tserkovnyak:2005_RMP,Hoffmann:2013_IEEE}]. In such a scenario, $V_\mathrm{ISHE}$ due to the inverse spin Hall effect cannot be detected in the NM. However, the asymmetric voltages can still be seen, being independent of spin pumping. In the present case of $n$-GaN:Si, the spin diffusion length is $\lambda_\mathrm{N}$ $\sim$ 80 nm. It was reported that for NM metal systems, the asymmetry voltage contribution for a film thickness of the order of $\lambda_\mathrm{N}$ is close to 100\% of the total emf measured. This is due to the dominance of $I_\mathrm{bf}$ over $I_\mathrm{sp}$ [\onlinecite{Vlaminck:2013_PRB}]. However, with a greater thickness of the NM film, the asymmetric contribution diminishes and approaches a minimum saturation value for a film thickness  $\gg\lambda_\mathrm{N}$. Furthermore, it was reported by Flovik $et\,al.$ \cite{Flovik:2015_JAP} that this strong $V_\mathrm{Asym}$ could be assigned to Eddy current effects and it was shown that in the case of a NM layer of Pt, the asymmetry decreases considerably with increasing the thickness of the NM layer. In the systems considered here, for a thin layer of $n$-GaN:Si the spin backflow is likely to be the dominant mechanism that leads to the suppression of the $V_\mathrm{sym}$ signal, since the Oersted fields induce a significant distortion of the FMR lineshape, not observed for the $1.9\,\mu\mathrm{m}$ thick $n$-GaN:Si sample. For the case of ZnO, also a wide band gap wurtzite semiconductor like GaN, D'Ambrosio $et\,al.$ [\onlinecite{D'Ambrosio:2015_JJAP}] reported the measured emf from the ZnO layer in a Py/ZnO bilayer under FMR conditions, as due to PHE in the Py layer. In our case of $n$-GaN:Si, the control experiments -- as previously mentioned --did not reveal any measurable voltage due to PHE in the Py layer. Moreover, the thickness of the ZnO film used in case of D'Ambrosio $et\,al.$ was 200 nm, making spin backflow a likely reason for the observed dominant contribution from PHE. On the other hand, in our case for calculating the spin Hall angle $\theta_\mathrm{SH}$ we carry out measurements on a $1.9\,\mu\mathrm{m}$ thick $n$-GaN:Si layer, which can be considered as a bulk system where the role of spin backflow is largely suppressed -- as discussed previously in detail -- upon comparison with the 150 nm $n$-GaN:Si layer. A detailed investigation of the dependence of spin pumping on the thickness and carrier concentration of $n$-GaN:Si is needed for an in-depth understanding of this behaviour.

\section{Summary}

We have provided experimental demonstration of spin pumping induced spin current generation and its detection at room temperature using the ISHE in the Rashba semiconductor $n$-GaN:Si. From the fundamental model of spin pumping, a spin mixing conductance of ($1.38\times10^{18})\,\mathrm{m}^{-2}$ for the Py/$n$-GaN:Si interface and a spin Hall angle $\theta_\mathrm{SH}=3.03\times10^{-3}$ for wz $n$-GaN:Si are found. The value obtained for $\theta_\mathrm{SH}$ is at least one order of magnitude higher than those of other semiconductors like Si, Ge, ZnO and $n$-GaAs. The experimental demonstration of generation of pure spin current in $n$-GaN:Si and its enhanced spin-charge conversion efficiency over other functional semiconductors points at III-nitrides as model systems for studies on spin-related phenomena in non-centrosymmetric semiconductors. Moreover, this work  paves the way to the realization of nitride-based low power optoelectronic, non-volatile and low dissipative spin devices like $e.g.$ spin batteries.

\section*{Acknowledgements}

This work was supported by the Austrian Science Foundation --
FWF (P22477 and P24471 and P26830), by the NATO Science for Peace Programme (Project No. 984735)
and by the EU $7^{\mathrm{th}}$  Framework Programmes:
CAPACITIES project REGPOT-CT-2013-316014 (EAgLE), EIT+ (NanoMat P2IG.01.01.02-02-002/08) and
FunDMS Advanced Grant of the European Research Council (ERC grant No.\,227690).

\bibliographystyle{apsrev4-1}

%

\end{document}